\documentclass[10pt,showpacs,amsmath,amssymb]{article}
\usepackage{amsfonts}
\usepackage{mathdots}
\usepackage{color}
\usepackage{graphicx}
\begin{document}

\title{A $N=2$ extension of the Hirota bilinear formalism and the supersymmetric KdV equation}
\author{Laurent Delisle}

\author{Laurent Delisle${}^{1,2}$}

\footnotetext[1]{Institut de math\'ematiques de Jussieu-Paris Rive Gauche, UP7D-Campus des Grands Moulins, B\^atiment Sophie Germain, Cases 7012, 75205 Paris Cedex 13.}
\footnotetext[2]{email:laurent.delisle@imj-prg.fr}

\date{\today}

\maketitle

\begin{abstract}
We present a bilinear Hirota representation of the $N=2$ supersymmetric extension of the Korteweg-de Vries equation. This representation is deduced using binary Bell polynomials, hierarchies and fermionic limits. We, also, propose a new approach for the generalisation of the Hirota bilinear formalism in the $N=2$ supersymmetric context.
\end{abstract}


\section{Introduction}

The study of exact solutions of completely integrable supersymmetric systems is of current interest in modern mathematical physics research. In particular, the $N=2$ supersymmetric extension of the Korteweg-de Vries (KdV) equation \cite{labelle} has been largely studied in terms of integrability conditions, exact solutions and symmetry group structures \cite{ayari,ghosh,hussin,zhang,delisle,delisle1,hussin1,liu,tian}. The equation is described by a bosonic superfield $A$ defined on the superspace $\mathbb{R}^{2\vert 2}$ \cite{cornwell} of local coordinates $(x,t,\theta_1,\theta_2)$. The variables $(x,t)$ are the usual Euclidean space-time coordinates and $(\theta_1,\theta_2)$ are real Grassmann coordinates satisfying the usual anti-commutation relations
\begin{equation}
\theta_1\theta_2+\theta_2\theta_1=0,\quad \theta_1^2=\theta_2^2=0.
\end{equation}
The superfield $A$ satisfies the Labelle-Mathieu $N=2$ supersymmetric extension of the KdV equation \cite{labelle}
\begin{equation}
A_t=(-A_{xx}+(a+2)AD_1D_2A+(a-1)(D_1A)(D_2A)+aA^3)_x,
\label{kdv}
\end{equation}
where $a\in \mathbb{R}$ is a real parameter, $D_1$, $D_2$ are the covariant derivatives defined as
\begin{equation}
D_i=\partial_{\theta_i}+\theta_i\partial_x,\quad i=1,2
\end{equation}
and satisfy $D_1^2=D_2^2=\partial_x$. The bosonic superfield $A$ can be decomposed \cite{cornwell} using a Taylor expansion as
\begin{equation}
A(x,t,\theta_1,\theta_2)=u(x,t)+\theta_1\xi_1(x,t)+\theta_2\xi_2(x,t)-\theta_1\theta_2v(x,t),
\label{fieldA}
\end{equation}
where $u$, $v$ are complex-valued even functions and $\xi_1$, $\xi_2$ are complex-valued odd functions.

 Labelle and Mathieu \cite{labelle} showed that equation (\ref{kdv}) is completely integrable for the special choices $a=-2,1,4$. This fact suggests that, for these special values, the supersymmetric KdV equation possesses travelling wave and multi-soliton solutions \cite{ayari,ghosh,hussin,zhang,delisle,delisle1}. An algebraic direct method to find such solutions is described by the Hirota bilinear formalism. This method was used numerous time for non-supersymmetric integrable evolution equation to construct soliton and similarity solutions, B\"acklund and Darboux transformations and to obtain integrability conditions \cite{ablowitz}. Carstea has adapted this formalism to $N=1$ supersymmetric extensions \cite{carstea} such as the KdV, modified KdV and Sine-Gordon equations. The generalisation of this formalism to $N=2$ extensions has been confronted to numerous difficulties. Zhang and \textit{al.} \cite{zhang} used the strategy of decomposing equation (\ref{kdv}) into two $N=1$ equations for which the Hirota formalism is well adapted. The way of achieving this is to re-write the bosonic superfield $A$ given in (\ref{fieldA}) as
\begin{equation}
A(x,t,\theta_1,\theta_2)=A_0(x,t,\theta_1)+\theta_2\Xi(x,t,\theta_1),
\label{fieldAdec}
\end{equation}
where $A_0(x,t,\theta_1)=u(x,t)+\theta_1\xi_1(x,t)$ and $\Xi(x,t,\theta_1)=\xi_2(x,t)+\theta_1v(x,t)$ are, respectively, bosonic and fermionic superfields of $(x,t,\theta_1)\in \mathbb{R}^{2\vert 1}$. In order to use the bilinear Hirota formalism \cite{ablowitz}, we have to re-write the superfields $A_0$ and $\Xi$ in terms of dimensionless bosonic superfields. This is done by dimensional analysis following the fact that equation (\ref{kdv}) is invariant under the dilatation vector field \cite{ayari}
\begin{equation}
x\partial_x+3t\partial_t+\frac12\theta_1\partial_{\theta_1}+\frac12\theta_2\partial_{\theta_2}-A\partial_A.
\end{equation}
This vector field shows that, under the transformation
\begin{equation}
(x,t,\theta_1,\theta_2,A)\quad \longrightarrow\quad \left(\lambda x,\lambda^3t,\lambda^{\frac12}\theta_1,\lambda^{\frac12}\theta_2,\lambda^{-1}A\right),
\end{equation}
equation (\ref{kdv}) is invariant, where $\lambda$ is a free parameter. Under these transformations, we deduce the dimension of these quantities
\begin{equation}
[A]=-1,\quad [\partial_x]=-1,\quad [D_1]=-\frac12
\end{equation}
and this allows us to re-write the superfields $A_0$ and $\Xi$ as
\begin{equation}
A_0=\partial_x B,\quad \Xi=D_1p_x,
\label{dimfield}
\end{equation}
where $B=B(x,t,\theta_1)$ and $p(x,t,\theta_1)$ are bosonic superfields such that $[B]=[p]=0$. Introducing the superfield $A$, given by (\ref{fieldAdec}) together with the relations (\ref{dimfield}), in equation (\ref{kdv}), we get two $N=1$ supersymmetric equations
\begin{eqnarray}
B_t&=&-B_{xxx}+(a+2)B_xp_{xx}+(a-1)(D_1B_x)(D_1p_x)+aB_x^3,\label{kdvN11}\\
D_1p_t&=&-D_1p_{xxx}+3p_{xx}D_1p_x-(a+2)B_xD_1B_{xx}+(1-a)B_{xx}D_1B_x+3aB_x^2D_1p_x.\label{kdvN12}
\end{eqnarray}
The study of these two equations is the main object of the paper for the special cases $a=-2,1,4$, \textit{i.e.} for the cases for which equation (\ref{kdv}) is completely integrable \cite{labelle}. Indeed, we will give a bilinear Hirota representation of these equations using different approaches using the binary Bell polynomials \cite{fan}, hierarchies \cite{hussin1,liu,tian} and fermionic limits \cite{hussin}. The later approach is simple, it relies on taking $\xi_1=\xi_2=0$ in the representation of the bosonic superfield $A$ given in (\ref{fieldA}). The binary Bell polynomials have, recently, found a connection with the Hirota bilinear formalism \cite{fan} and this connection will be used throughout this paper. 

The one-variable Bell polynomials $Y$ are defined as
\begin{equation}
Y_{k_x x,k_tt,k_1\theta_1}(f)=e^{-f}\partial_x^{k_x}\partial_t^{k_t}D_1^{k_1}e^f,
\end{equation}
where $k_{\mu}$ are integer constants and $f=f(x,t,\theta_1)$ is a bosonic superfield. Using the polynomials $Y$, we define the binary Bell polynomials $\mathcal{Y}$ as
\begin{equation}
\mathcal{Y}_{k_xx,k_tt,k_1\theta_1}(w_1,w_2)=Y_{k_xx,k_tt,k_1\theta_1}(f_{\tilde{k}_xx\tilde{k}_tt\tilde{k}_1\theta_1}),
\label{binBell}
\end{equation}
where the different derivatives of $f$ are replaced by the superfields $w_1$ and $w_2$ following the procedure
\begin{equation}
 f_{\tilde{k}_xx\tilde{k}_tt\tilde{k}_1\theta_1}=\left\{
	\begin{array}{ll}
		w_{1,\tilde{k}_x x\tilde{k}_tt\tilde{k}_1\theta_1}  & \mbox{if } \tilde{k}_x+\tilde{k}_t+\tilde{k}_1 \,\,\mbox{is odd}, \\
		w_{2,\tilde{k}_x x\tilde{k}_tt\tilde{k}_1\theta_1} & \mbox{if } \tilde{k}_x+\tilde{k}_t+\tilde{k}_1\,\, \mbox{is even}.
	\end{array}
	\right.
\end{equation}
Note here that we are using the notation $f_{\tilde{k}_xx\tilde{k}_tt\tilde{k}_1\theta_1}=\partial_x^{\tilde{k}_x}\partial_t^{\tilde{k}_t}D_1^{\tilde{k}_1}f$. The link with the Hirota bilinear formalism is given by
\begin{equation}
\mathcal{Y}_{k_xx,k_tt,k_1\theta_1}\left(w_1=\ln\left(\frac{f}{g}\right),w_2=\ln(fg)\right)=(fg)^{-1}\mathcal{S}_1^{k_1}\mathcal{D}_x^{k_x}\mathcal{D}_t^{k_t}(f\cdot g),
\label{link}
\end{equation}
where the Hirota derivative is defined as
\begin{equation}
\mathcal{S}_1^{k_1}\mathcal{D}_x^{k_x}\mathcal{D}_t^{k_t}(f\cdot g)=(D_1-D_1^{\prime})^{k_1}(\partial_x-\partial_{x^{\prime}})^{k_x}(\partial_t-\partial_{t^{\prime}})^{k_t}f(x,t,\theta_1)g(x^{\prime},t^{\prime},\theta_1^{\prime})\vert_{(x^{\prime},t^{\prime},\theta_1^{\prime})=(x,t,\theta_1)}.
\end{equation}
These relations will be used to transform equations (\ref{kdvN11}) and (\ref{kdvN12}) into binary Bell polynomial equations. To achieve this, we will need auxiliary tools such as hierarchies of the supersymmetric KdV equation (\ref{kdv}) and fermionic limits.

This paper is divided as follows. In the following three sections, we give a bilinear representation of the supersymmetric KdV equation (\ref{kdv}) for, respectively, $a=1$, $a=4$ and $a=-2$. In section 2, we use directly the binary Bell polynomials to get a general Hirota formulation. In the $a=4$ case, we use the binary Bell polynomials and the Two-Boson supersymmetric equation \cite{zhang,hussin1}, which are members of the same hierarchy, to obtain a Hirota equation. In section 4, we get the bilinear representation using fermionic limits and retrieve the bosonic Miura transformation \cite{ablowitz} linking a solution of the KdV equation with the modified KdV equation. The last section addresses the open problem of generalizing the Hirota bilinear formalism to the $N=2$ supersymmetric context. We exhibit this generalisation threw the supersymmetric KdV equation with $a=1$.  We conclude the paper with some future outlooks and remarks.

The novelty of this paper is based on the use of the binary Bell polynomials (\ref{binBell}) to get a bilinear representation of the supersymmetric KdV equation (\ref{kdv}). As of today, the Hirota formulation of the supersymmetric KdV equation with $a=-2$ was an open problem and here we propose a partial answer to this question. We also, for the first time, generalize the Hirota bilinear formulation to $N=2$ extensions of certain integrable systems.

\section{The supersymmetric KdV equation with $a=1$}
In this section, we directly use the binary Bell polynomials to get a bilinear representation of the supersymmetric KdV equation (\ref{kdv}) with $a=1$. In this case, the two $N=1$ equations (\ref{kdvN11}) and (\ref{kdvN12}) reduces to
\begin{eqnarray}
B_t&=&-B_{xxx}+3B_xp_{xx}+B_x^3,\label{kdva11}\\
D_1p_t&=&-D_1p_{xxx}+3p_{xx}D_1p_{x}-3B_xD_1B_{xx}+3B_x^2D_1p_x.\label{kdva12}
\end{eqnarray}
We can notice from these equations that the superfield $B$ is associated to odd numbers of derivatives while the superfield $p$ to even numbers. This fact is compatible with the binary Bell polynomials \cite{fan}. Indeed, we have the expressions
\begin{eqnarray}
\mathcal{Y}_{t}(cB,dp)=cB_t,\quad \mathcal{Y}_{3x}(cB,dp)=cB_{xxx}+3cd B_xp_{xx}+c^3B_x^3,
\end{eqnarray}
from which we easily deduce an equivalent representation of equation (\ref{kdva11}) given as
\begin{equation}
\mathcal{Y}_t(iB,-p)+\mathcal{Y}_{3x}(iB,-p)=0.
\end{equation} 
For the second equation (\ref{kdva12}), we have the following binary Bell polynomials, assuming equation (\ref{kdva11}) is satisfied,
\begin{eqnarray}
\mathcal{Y}_{t\theta_1}(iB,-p)&=&-D_1p_t+B_{xxx}D_1B-3B_xp_{xx}D_1B-B_x^3D_1B,\\
\mathcal{Y}_{xxx\theta_1}(iB,-p)&=&-D_1p_{xxx}+B_x^3D_1B+3B_xp_{xx}D_1B-B_{xxx}D_1B+3B_x^2D_1p_x\nonumber\\
&+&3p_{xx}D_1p_x-3B_xD_1B_{xx}
\end{eqnarray}
and it is direct to show that equation (\ref{kdva12}) is equivalent to the binary Bell polynomials equation
\begin{equation}
\mathcal{Y}_{t\theta_1}(iB,-p)+\mathcal{Y}_{xxx\theta_1}(iB,-p)=0.
\end{equation}
Using the link between the binary Bell polynomials and the Hirota derivative (\ref{link}), we obtain, casting the change of variables $iB=\ln(f/g)$ and $-p=\ln(fg)$, the bilinear representation of the supersymmetric KdV equation with $a=1$ \cite{zhang} given as
\begin{equation}
(\mathcal{D}_t+\mathcal{D}_x^3)(f\cdot g)=0,\quad \mathcal{S}_1(\mathcal{D}_t+\mathcal{D}_x^3)(f\cdot g)=0.
\end{equation}

\section{The supersymmetric KdV equation with $a=4$}
In this case, we give a Bell polynomial perspective to the supersymmetric KdV equation with $a=4$ using its integrable hierarchy \cite{hussin1,tian}. Unlike the $a=1$ case, the Bell polynomial approach may not be directly applied to equation (\ref{kdvN11}) and (\ref{kdvN12}) with $a=4$. Indeed, these equations are explicitly given as
\begin{eqnarray}
B_t&=&-B_{xxx}+6B_xp_{xx}+3(D_1B_x)(D_1p_x)+4B_x^3,\label{kdva41}\\
D_1p_t&=&-D_1p_{xxx}+3p_{xx}D_1p_x-6B_xD_1B_{xx}-3B_{xx}D_1B_x+12B_x^2D_1p_x,\label{kdva42}
\end{eqnarray}
and the terms $(D_1B_x)(D_1p_x)$, $B_{xx}D_1B_x$ are incompatible with the definition of the binary Bell polynomials. These terms are incompatible in the sense that the superfield $B$ is associated to odd number of derivatives in the binary Bell polynomial perspective. So, we have to find an other way of writing these problematic terms. One way of achieving this is by considering the integrable hierarchy associated to the supersymmetric KdV equation with $a=4$. One member of this hierarchy is the Two-Boson supersymmetric equation \cite{zhang,hussin1} for which its flow commutes with the flow associated to equation (\ref{kdv}) with $a=4$. Regarding this matter, the Two-Boson supersymmetric equation is given by
\begin{eqnarray}
w_{t_2}&=&(-w_x+w^2+2D_1\rho)_x,\label{tb1}\\
\rho_{t_2}&=&(\rho_x+2w\rho)_x,\label{tb2}
\end{eqnarray}
where $w$ and $\rho$ are, respectively, bosonic and fermionic superfields. Making use of the dilatation invariant vector field, we cast the change of variables $w=C_x$ and $\rho=D_1q_x$, where $C$ and $q$ are dimensionless bosonic superfields. In this case, after integration with respect to the variable $x$, equations (\ref{tb1}) and (\ref{tb2}) reduce to
\begin{eqnarray}
C_{t_2}&=&-C_{xx}+C_x^2+2q_{xx},\\
D_1q_{t_2}&=&D_1q_{xx}+2C_xD_1q_{x}.
\end{eqnarray}
At first sight, this system is again incompatible with the binary Bell polynomials, but, taking $m=2q-C$ and $n=C$, these equations are given as
\begin{eqnarray}
n_{t_2}&=&m_{xx}+n_x^2,\label{tb11}\\
D_1m_{t_2}&=&D_1n_{xx}+2n_xD_1m_x.\label{tb12}
\end{eqnarray}
We thus observe that $m$ is associated to even numbers of derivatives, while $n$ to odd numbers. From the binary Bell polynomials
\begin{eqnarray}
\mathcal{Y}_{t_2}(n,m)&=&n_{t_2},\quad \mathcal{Y}_{xx\theta_1}(n,m)=m_{xx}D_1n+n_x^2D_1n+D_1n_{xx}+2n_xD_1m_x \\
\mathcal{Y}_{xx}(n,m)&=&m_{xx}+n_x^2,\quad \mathcal{Y}_{t_2\theta_1}(n,m)=D_1m_{t_2}+m_{xx}D_1n+n_x^2D_1n,
\end{eqnarray}
it is direct to see that the system of equations (\ref{tb11}) and (\ref{tb12}) is equivalent to the system of binary Bell polynomials
\begin{equation}
\mathcal{Y}_{t_2}(n,m)-\mathcal{Y}_{xx}(n,m)=0,\quad \mathcal{Y}_{t_2\theta_1}(n,m)-\mathcal{Y}_{xx\theta_1}(n,m)=0.
\end{equation}
Making the change of variables $n=\ln(f/g)$, $m=\ln(fg)$ and using the link between the Bell polynomials and the Hirota derivative (\ref{link}), we get the bilinear representation of the Two-Boson equation \cite{liu1} given as
\begin{equation}
(\mathcal{D}_{t_2}-\mathcal{D}_{x}^2)(f\cdot g)=0,\quad \mathcal{S}_1(\mathcal{D}_{t_2}-\mathcal{D}_x^2)(f\cdot g)=0.
\end{equation}
Let us now focus our attention on the first equation (\ref{kdva41}) of the supersymmetric KdV equation for $a=4$. We have, taking $n=\alpha B$ and $m=\beta p$, the following binary Bell polynomial expression:
\begin{equation}
\mathcal{Y}_{t_2x}(n,m)=\mathcal{Y}_{t_2x}(\alpha B,\beta p)=\alpha B_{xxx}+3\alpha\beta B_xp_{xx}+\alpha^3 B_x^3+2\alpha\beta(D_1B_x)(D_1p_x),
\end{equation}
where we have retrieve the previously problematic term $(D_1B_x)(D_1p_x)$. Supposing that equation (\ref{kdva41}) has the binary Bell polynomial representation
\begin{equation}
\frac{1}{\alpha}\mathcal{Y}_t(\alpha B,\beta p)=\gamma \mathcal{Y}_{xxx}(\alpha B,\beta p)+\delta \mathcal{Y}_{t_2x}(\alpha B,\beta p),
\end{equation}
for $\alpha,\beta,\gamma,\delta\in\mathbb{C}$ constants, we directly find that
\begin{equation}
\alpha=2i,\quad \beta=-2,\quad \gamma=\frac{i}{8},\quad \delta=\frac{3i}{8}
\end{equation}
and, thus, equation (\ref{kdva41}) as the following Hirota bilinear formulation, taking $2i B=\ln(f/g)$ and $-2p=\ln(fg)$,
\begin{equation}
\left(\mathcal{D}_t+\frac14\mathcal{D}_x^3+\frac34\mathcal{D}_x\mathcal{D}_{t_2}\right)(f\cdot g)=0.
\label{bilia41}
\end{equation}
The Bell polynomial analysis of the second equation (\ref{kdva42}) is similar as for the first one. Indeed, it lies on the following binary Bell polynomial expressions:
\begin{eqnarray*}
\frac{1}{\beta}\mathcal{Y}_{t\theta_1}(\alpha B,\beta p)&=&D_1p_t-2B_{xxx}D_1B+12B_xp_{xx}D_1B+6(D_1B_x)(D_1p_x)D_1B\\&+&8B_x^3D_1B,\\
\frac{1}{\beta}\mathcal{Y}_{xt_2\theta_1}(\alpha B, \beta p)&=&2B_{xxx}D_1B-8(D_1B_x)(D_1p_x)D_1B-12B_xp_{xx}D_1B-8B_x^3D_1B\\
&+&D_1p_{xxx}+4B_{xx}D_1B_x+6B_xD_1B_{xx}-12B_x^2D_1p_x-2p_{xx}D_1p_x,\\
\frac{1}{\beta}\mathcal{Y}_{xxx\theta_1}(\alpha B,\beta p)&=&2B_{xxx}D_1B-12B_xp_{xx}D_1B-8B_x^3D_1B+D_1p_{xxx}-6p_{xx}D_1p_x\\
&+&6B_xD_1B_{xx}-12B_x^2D_1p_x.
\end{eqnarray*}
From these expressions, it is elementary algebra to show that equation (\ref{kdva42}) is equivalent to the binary Bell polynomials equation
\begin{equation}
\mathcal{Y}_{t\theta_1}(2iB,-2p)+\frac14\mathcal{Y}_{xxx\theta_1}(2iB, -2p)+\frac34\mathcal{Y}_{xt_2\theta_1}(2iB,-p)=0
\end{equation}
and, using the same change of variables for $B$ and $p$ as for equation (\ref{kdva41}), we get the bilinear representation
\begin{equation}
\mathcal{S}_1\left(\mathcal{D}_t+\frac14\mathcal{D}_x^3+\frac34\mathcal{D}_{x}\mathcal{D}_{t_2}\right)(f\cdot g)=0.
\label{bilia42}
\end{equation}
Hence, equations (\ref{bilia41}) and (\ref{bilia42}) represent the Hirota formulation of the supersymmetric KdV equation with $a=4$ \cite{zhang}, where the functions $B$, $p$, $f$ and $g$ are related as
\begin{equation}
B=-\frac{i}{2}\ln\left(\frac{f}{g}\right),\quad p=-\frac{1}{2}\ln(fg).
\end{equation}

\section{The supersymmetric KdV equation with $a=-2$}

As of today, a Hirota bilinear representation of the supersymmetric KdV equation with $a=-2$ was an open problem \cite{zhang}. In this section, we propose a partial answer to this problem by considering the fermionic limit \cite{hussin} of the superfield $A$ given in (\ref{fieldA}). This means that we take $\xi_1=\xi_2=0$ in (\ref{fieldA}) and we consider bosonic superfield $A$ of the form
\begin{equation}
A(x,t,\theta_1,\theta_2)=u(x,t)-\theta_1\theta_2v(x,t)
\end{equation}
solution of equation (\ref{kdv}) with $a=-2$. In this case, the bosonic complex-valued functions $u$ and $v$ satisfy the system of partial differential equations
\begin{eqnarray}
u_t&=&-u_{xxx}-6u^2u_x,\label{kdva21}\\
v_t&=&-v_{xxx}+6vv_x+6u_xu_{xx}-6u^2v_x-12uu_xv.\label{kdva22}
\end{eqnarray}
We can make some observations on this system: equation (\ref{kdva21}) is the bosonic modified KdV equation for which a Hirota formulation is known and, as a second observation, we have that, taking $u=0$, this system reduces to the bosonic KdV equation \cite{ablowitz}. Again, at first sight, this system is incompatible with the definition of the binary Bell polynomials. To solve this problem, we use a Miura-type transformation relating the supersymmetric KdV equation with $a=-2$ to the equation \cite{tian}
\begin{eqnarray}
Q_t&=&-Q_{xxx}+\frac12 Q_x^3+\frac34 (D_2Q_x)(D_2Q)Q_x-\frac34 (D_2Q_x)(D_1Q)(D_1D_2Q)\nonumber\\
&-&\frac34(D_2Q)(D_1Q_x)(D_1D_2Q)+\frac{3}{4}(D_1Q_x)(D_1Q)Q_x,
\label{equaQ}
\end{eqnarray}
where $Q$ is a bosonic superfield. The equation (\ref{equaQ}) is the first non-trivial flow of a $N=2$ supersymmetric hierarchy, as shown by Tian and Liu \cite{tian}, and the Miura-type transformation is explicitly given as
\begin{equation}
A=\frac12 D_1D_2Q+\frac14 (D_2Q)(D_1Q).\label{miura}
\end{equation}
Using the Taylor expansion $Q(x,t,\theta_1,\theta_2)=q_0(x,t)-\theta_1\theta_2q_{12}(x,t)$ (fermionic limit), the Miura-type transformation (\ref{miura}) is equivalent, in components, to
\begin{equation}
u=\frac12 q_{12},\quad v=\frac{1}{4}(q_{12}^2+q_{0,x}^2)-\frac12 q_{0,xx}.
\end{equation}
Note that setting $u=0$ in these transformations leads to the classical Miura transformation $v=\frac14 q_{0,x}^2-\frac12 q_{0,xx}$ relating a solution of the modified KdV equation to a solution of the KdV equation \cite{ablowitz}. This observation is compatible with the fact that the bosonic complex-valued functions $q_0$ and $q_{12}$ satisfy the decoupled system of partial differential equations
\begin{eqnarray}
q_{0,t}&=&-q_{0,xxx}+\frac12 q_{0,x}^3,\\
q_{12,t}&=&-q_{12,xxx}-\frac32 q_{12}^2q_{12,x}.
\end{eqnarray}
These two equations are all of the modified KdV-type and we know that they possess a Hirota bilinear representation \cite{ablowitz}. Indeed, they are equivalent to the binary Bell polynomial equations
\begin{eqnarray}
\mathcal{Y}_t(1/2q_0,\beta \tilde{q}_0)+\mathcal{Y}_{xxx}(1/2q_0,\beta\tilde{q}_0)-3\mathcal{Y}_x(1/2q_0,\beta \tilde{q}_0)\mathcal{Y}_{xx}(1/2q_0,\beta\tilde{q}_0)&=&0,\\
\mathcal{Y}_t(i/2p_{12},\gamma\tilde{p}_{12})+\mathcal{Y}_{xxx}(i/2p_{12},\gamma\tilde{p}_{12})-3\mathcal{Y}_x(i/2p_{12},\gamma\tilde{p}_{12})\mathcal{Y}_{xx}(i/2p_{12},\gamma \tilde{p}_{12})&=&0,
\end{eqnarray}
where $q_{12}=p_{12,x}$, $\beta$ and $\gamma$ are arbitrary constants and $\tilde{q}_0$ and $\tilde{p}_{12}$ are auxiliary and arbitrary bosonic functions. Using the change of variables
\begin{equation}
\frac12 q_0=\ln\left(\frac{f}{g}\right),\quad \beta\tilde{q}_0=\ln(fg),\quad \frac{i}{2}p_{12}=\ln\left(\frac{\tilde{f}}{\tilde{g}}\right),\quad \gamma \tilde{p}_{12}=\ln(\tilde{f}\tilde{g}),
\end{equation}
we get the Hirota bilinear representation
\begin{eqnarray}
(\mathcal{D}_t+\mathcal{D}_x^3)(f\cdot g)&=&3\lambda\mathcal{D}_x(f\cdot g),\quad \mathcal{D}_x^2(f\cdot g)=\lambda fg,\label{bilia21}\\ (\mathcal{D}_t+\mathcal{D}_x^3)(\tilde{f}\cdot \tilde{g})&=&3\tilde{\lambda}\mathcal{D}_x(\tilde{f}\cdot \tilde{g}),\quad \mathcal{D}_x^2(\tilde{f}\cdot \tilde{g})=\tilde{\lambda} \tilde{f}\tilde{g},\label{bilia22}
\end{eqnarray}
where $\lambda$ and $\tilde{\lambda}$ are arbitrary parameters and the functions $u$ and $v$ take the explicit forms
\begin{equation}
u=-i\partial_x\ln\left(\frac{\tilde{f}}{\tilde{g}}\right),\quad v=(\partial_x\ln f/g)^2-(\partial_x\ln\tilde{f}/\tilde{g})^2-\partial_x^2\ln f/g.
\end{equation}

In the reminder of this section, we give plots of different soliton solutions. For a $1$-soliton profile, the functions $f_1$, $g_1$, $\tilde{f}_1$ and $\tilde{g}_1$, solution of the bilinear equations (\ref{bilia21}) and (\ref{bilia22}) with $\lambda=\tilde{\lambda}=0$, may be chosen as
\begin{equation}
f_1=1+e^{\eta},\quad g_1=1-e^{\eta},\quad \tilde{f}_1=1+ie^{\tilde{\eta}},\quad \tilde{g}_1=1-ie^{\tilde{\eta}},
\end{equation} 
where $\eta=\kappa x-\kappa^3 t$ and $\tilde{\eta}=\tilde{\kappa}x-\tilde{\kappa}^3t$. In the case of a $2$-soliton profile, the functions $f_2$, $g_2$, $\tilde{f}_2$ and $\tilde{g}_2$ given as
\begin{eqnarray}
f_2&=&1+e^{\eta_1}+e^{\eta_2}+A_{12}e^{\eta_1+\eta_2},\quad g_2=1-e^{\eta_1}-e^{\eta_2}+A_{12}e^{\eta_1+\eta_2},\\
\tilde{f}_2&=&1+i e^{\tilde{\eta}_1}+ie^{\tilde{\eta}_2}-\tilde{A}_{12}e^{\tilde{\eta}_1+\tilde{\eta}_2},\quad \tilde{g}_2=1-i e^{\tilde{\eta}_1}-ie^{\tilde{\eta}_2}-\tilde{A}_{12}e^{\tilde{\eta}_1+\tilde{\eta}_2}
\end{eqnarray}
solves the bilinear equations (\ref{bilia21}) and (\ref{bilia22}) for $\lambda=\tilde{\lambda}=0$, where $\eta_i=\kappa_i x-\kappa_i^3t$, $\tilde{\eta}_i=\tilde{\kappa}_i x-\tilde{\kappa}_i^3t$ for $i=1,2$, $A_{12}=\left(\frac{\kappa_1-\kappa_2}{\kappa_1+\kappa_2}\right)^2$ and $\tilde{A}_{12}=\left(\frac{\tilde{\kappa}_1-\tilde{\kappa}_2}{\tilde{\kappa}_1+\tilde{\kappa}_2}\right)^2$. In the figures below, we use the notation
\begin{equation}
u_{(m,n)}=-i\partial_x\ln\left(\frac{\tilde{f}_n}{\tilde{g}_n}\right),\quad v_{(m,n)}=(\partial_x\ln f_m/g_m)^2-(\partial_x\ln \tilde{f}_n/\tilde{g}_n)^2-\partial_x^2\ln f_m/g_m
\end{equation}
and we have made the choices $\kappa=1$, $\tilde{\kappa}=\frac45$, $\kappa_1=\frac35$, $\kappa_2=\frac12$, $\tilde{\kappa}_1=\frac34$ and $\tilde{\kappa}_2=\frac23$. Figure 1 represents the functions $u_{(1,1)}$ and $u_{(2,2)}$, figure 2 gives the behavior of the functions $v_{(1,1)}$ and $v_{(2,2)}$, while figure 3 plots the functions $v_{(1,2)}$ and $v_{(2,1)}$.
\begin{figure}[h!]
\centering
\includegraphics[width=2in]{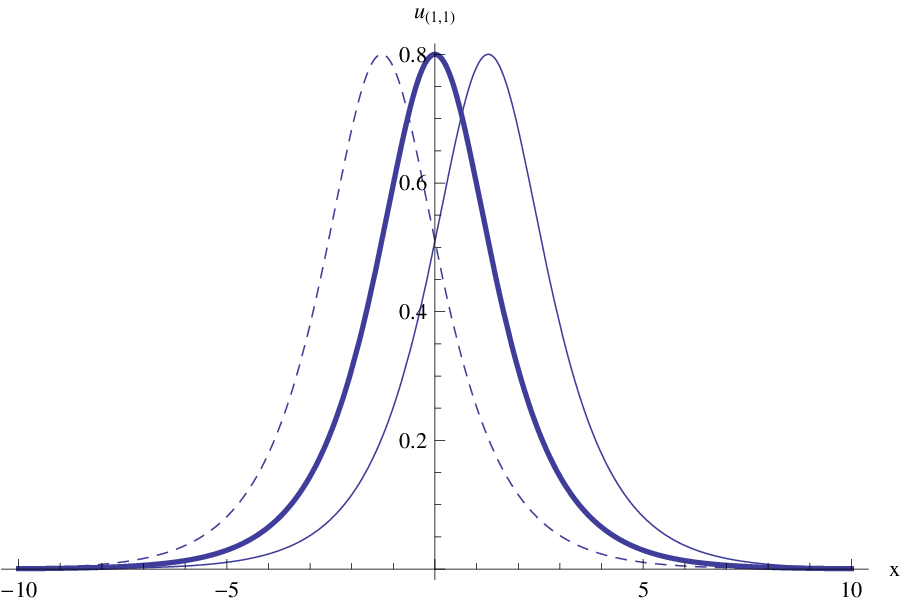}
\includegraphics[width=2in]{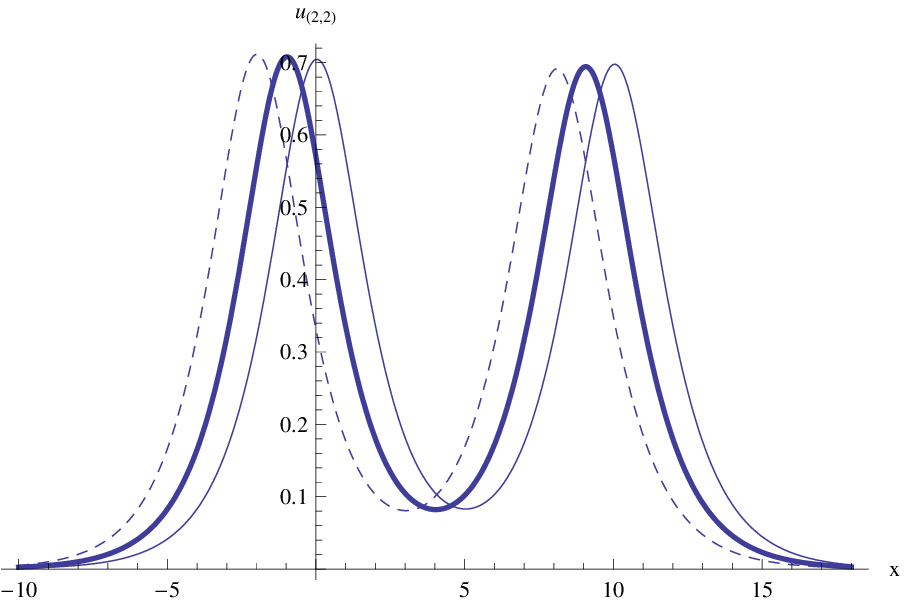}
\caption{The plot of the functions $u_{(1,1)}$ and $u_{(2,2)}$. The dashed curve correspond to time $t=-2$, the thick curve to time $t=0$ and the other curve to time $t=2$. }
\end{figure}
\begin{figure}[h!]
\centering
\includegraphics[width=2in]{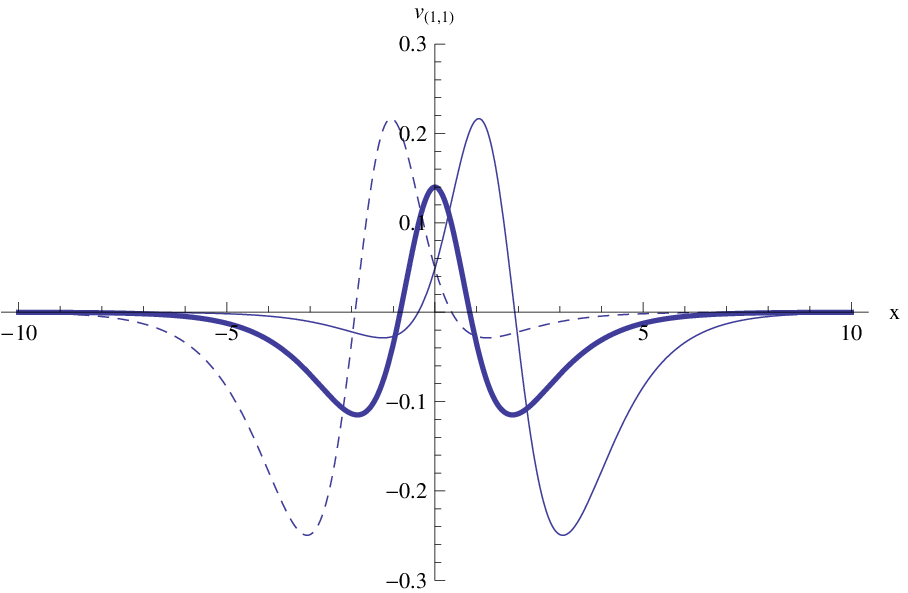}
\includegraphics[width=2in]{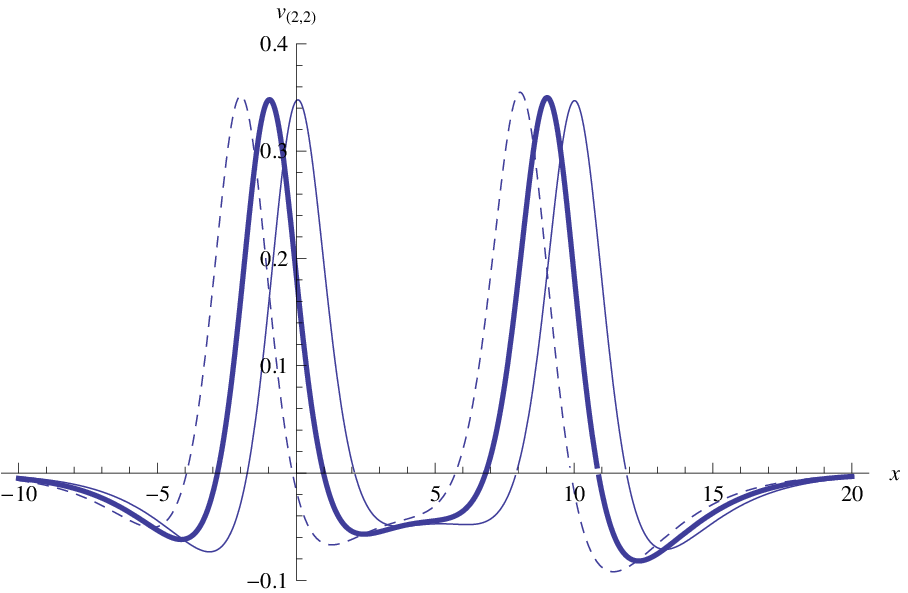}
\caption{The plot of the functions $v_{(1,1)}$ and $v_{(2,2)}$. The dashed curve correspond to time $t=-2$, the thick curve to time $t=0$ and the other curve to time $t=2$.}
\end{figure}
\begin{figure}[h!]
\centering
\includegraphics[width=2in]{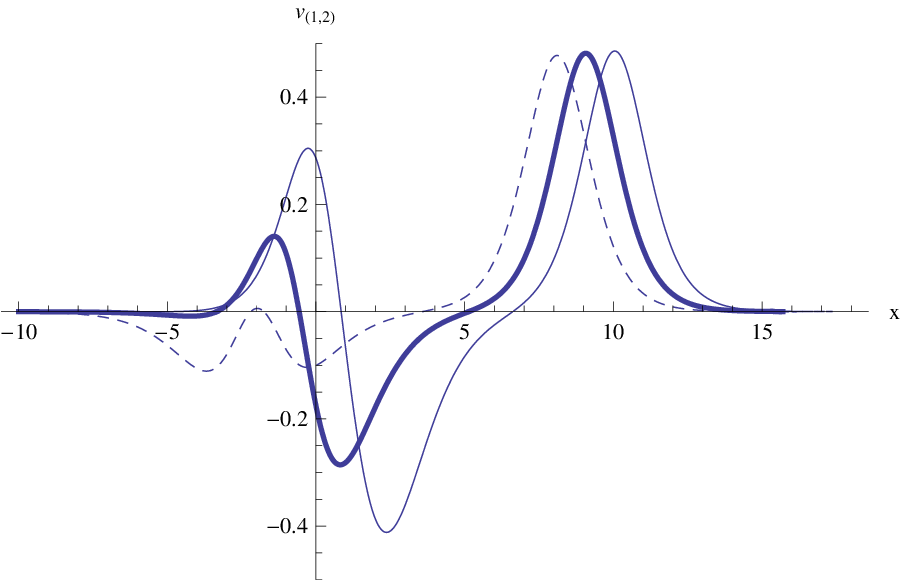}
\includegraphics[width=2in]{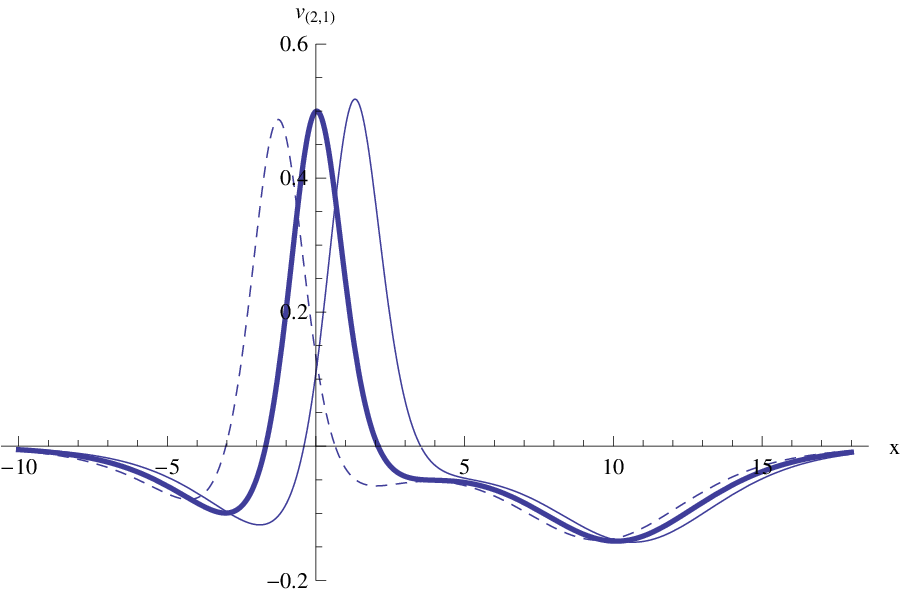}
\caption{The plot of the functions $v_{(1,2)}$ and $v_{(2,1)}$. The dashed curve correspond to time $t=-2$, the thick curve to time $t=0$ and the other curve to time $t=2$. }
\end{figure}

\section{The $N=2$ formalism: a first approach}

In this section, we make an attempt to formalize the Hirota bilinear approach in the $N=2$ supersymmetric context using the Bell polynomials. We will illustrate this new approach on the supersymmetric KdV equation (\ref{kdv}) with $a=1$. Using the change of variable $A=M_x$ and integrating once with respect to the variable $x$, the supersymmetric KdV equation for $a=1$ reads as
\begin{equation}
M_t=-M_{xxx}+3M_xD_1D_2M_x+M_x^3,
\label{kdvN2}
\end{equation}
where $M=M(x,t,\theta_1,\theta_2)$ is a dimensionless bosonic superfield. In order to adapt the Hirota bilinear formalism in the $N=2$ context, we make the following observation 
\begin{equation}
D_i^{-1}=\partial_x^{-1}D_i=\partial_x^{-1}\partial_{\theta_i}+\theta_i\quad\Longrightarrow\quad  D_1D_2M_x=\partial_x^2(D_1^{-1}D_2^{-1}M_x)=C_{xx}, 
\end{equation}
where $C=D_1^{-1}D_2^{-1}M_x$ and $\partial_x^{-1}$ is the integration operator with respect to $x$. In this setting, the equation (\ref{kdvN2}) can be re-written as
\begin{equation}
M_t=-M_{xxx}+3M_xC_{xx}+M_x^3
\end{equation}
and we observe that the Bell polynomial approach can be directly used. Indeed, the above equation is equivalent to the binary Bell polynomial equation
\begin{equation}
\mathcal{Y}_t(iM,-C)+\mathcal{Y}_{xxx}(iM,-C)=0
\end{equation}
which, using the identification $iM=\ln (f/g) $ and $-C=\ln(fg)$, can be written as
\begin{equation}
(\mathcal{D}_t+\mathcal{D}_{x}^3)(f\cdot g)=0,
\label{bilinearN2}
\end{equation}
where $f=f(x,t,\theta_1,\theta_2)$ and $g=g(x,t,\theta_1,\theta_2)$ are bosonic superfields. The identification imposes a further constraint given as
\begin{equation}
D_2D_1\ln\left(fg\right)=i\partial_x\ln\left(\frac{f}{g}\right)
\label{const}
\end{equation}
and it can be shown that this relation is equivalent to the bilinear Hirota equations
\begin{equation}
\mathcal{S}_2\mathcal{S}_1(f\cdot f)=2iff_x,\quad \mathcal{S}_2\mathcal{S}_1(g\cdot g)=-2igg_x.
\label{biliconst}
\end{equation}
In order to study this additional constraint, we consider the following Taylor expansions
\begin{equation}
\ln(f)=f_0+\theta_1f_1+\theta_2f_2+\theta_1\theta_2f_{12},\quad \ln(g)=g_0+\theta_1g_1+\theta_2g_2+\theta_1\theta_2g_{12}
\end{equation}
and, once introduce in (\ref{const}), we get
\begin{equation}
\ln(f)=f_0+(\theta_1-i\theta_2)f_1+i\theta_1\theta_2f_{0,x},\quad \ln(g)=g_0+(\theta_1+i\theta_2)g_1-i\theta_1\theta_2g_{0,x}.
\label{conscomp}
\end{equation}
For the $1$-soliton solution, we can choose 
\begin{equation}
f=1+a e^{\eta+\theta_1\zeta_1+\theta_2\zeta_2+\theta_1\theta_2m_{12}},\quad g=1+b e^{\eta+\theta_1\nu_1+\theta_2\nu_2+\theta_1\theta_2n_{12}}
\end{equation}
as a solution of the bilinear equation (\ref{bilinearN2}) for  $a,b,m_{12}, n_{12}$ bosonic quantities, $\zeta_1$, $\zeta_2$, $\nu_1$, $\nu_2$ fermionic quantities and $\eta=\kappa x-\kappa^3t$. These free parameters have to be determined in order that the constraint (\ref{const}) be satisfied. This can be done by substituting the above expressions for $f$ and $g$ in (\ref{conscomp}). We get the explicit forms
\begin{eqnarray}
f_0&=&\ln(1+ae^{\eta}),\quad f_1=a\zeta_1\frac{e^{\eta}}{1+ae^{\eta}},\quad \zeta_2=-i\zeta_1,\quad m_{12}=i\kappa,\\
g_0&=&\ln(1+be^{\eta}),\quad g_1=b\nu_1\frac{e^{\eta}}{1+be^{\eta}},\quad \nu_2=i\nu_1,\quad n_{12}=-i\kappa
\end{eqnarray}
and, as a consequence, the components of the superfield $A$ defined in (\ref{fieldA}) are given as
\begin{equation}
u=i(g_{0,x}-f_{0,x}),\quad \xi_1=i(g_{1,x}-f_{1,x}),\quad \xi_2=-(f_{1,x}+g_{1,x}),\quad v=-(f_{0,xx}+g_{0,xx}).
\end{equation}

In this section, we have produced, for the first time, a $N=2$ supersymmetric Hirota representation of the KdV equation with $a=1$. Indeed, the Hirota formulation is described by equations (\ref{bilinearN2}) and (\ref{biliconst}). We have thus partially solved the open problem of finding a supersymmetric $N=2$ generalisation of the Hirota bilinear formalism \cite{zhang}.

\section{Future outlooks and remarks}
In this paper, we have presented a systematic way of finding the Hirota bilinear representation of the $N=2$ supersymmetric KdV equation using its decomposition into two $N=1$ supersymmetric equations and the Bell polynomials. For the completely integrable cases $a=1,4$, we have obtain a complete representation using the integrable hierarchy and the binary Bell polynomials. For the $a=-2$ case, it was an open problem to find an Hirota formulation. Here, we have proposed a partial answer to this question using fermionic limits and have retrieved the well known Miura transformation relating the KdV and modified KdV equations. It still remains to find its general representation.

An other open problem was to find a $N=2$ generalisation of the Hirota bilinear formalism. We have, for the first time, succeeded in given such a representation for the supersymmetric KdV equation with $a=1$. The main idea in this construction was to re-write the operator $D_1D_2$ as a second derivative with respect to $x$ of a given quantity. This had the effect of transforming the $N=2$ equation into a "new" equation involving only derivatives of the bosonic variables $(x,t)$ for which the binary Bell polynomials could be directly applied. This as led to its $N=2$ Hirota representation.

Our future goals is to generalize the results of section 5. This new approach avoids treating a $N=2$ equation as two $N=1$ supersymmetric equations and this allows us to obtain a $N=2$ bilinear representation of the equation. As a final example to illustrate the efficiency of this new proposed procedure, we consider the $N=2$ supersymmetric extension of the potential Burgers equation
\begin{equation}
M_t=D_1D_2M_x+\frac12 M_x^2.
\end{equation}
As in section 5, we re-write the quantity $D_1D_2M_x$ as $C_{xx}$ where $C=D_1^{-1}D_2^{-1}M_x$ and, as a consequence, the Burgers equations \cite{hussin} reads as
\begin{equation}
M_t=C_{xx}+\frac{1}{2}M_x^2.
\label{Burgers}
\end{equation}
This equation may be cast into a binary Bell polynomials equation as
\begin{equation}
2\lambda \mathcal{Y}_t(\lambda M,2\lambda^2C)-\mathcal{Y}_{xx}(\lambda M,2\lambda^2C)=0,
\end{equation}
where $\lambda$ is a free complex parameter. Making the change of variables $\lambda M=\ln(f/g)$ and $2\lambda^2 C=\ln(fg)$, we get the Hirota bilinear equation
\begin{equation}
(2\lambda\mathcal{D}_t-\mathcal{D}_x^2)(f\cdot g)=0
\label{biliBurgers}
\end{equation}
together with the constraint
\begin{equation}
D_2D_1\ln(fg)=2\lambda\partial_x\ln\left(\frac{f}{g}\right),
\end{equation}
which can be re-write into a Hirota bilinear form as
\begin{equation}
\mathcal{S}_2\mathcal{S}_1(f\cdot f)=4\lambda ff_x,\quad \mathcal{S}_2\mathcal{S}_1(g\cdot g)=-4\lambda gg_x.
\label{biliconsBurgers}
\end{equation}
In conclusion, equations (\ref{Burgers}) and (\ref{biliconsBurgers}) constitute the Hirota bilinear representation of the supersymmetric potential Burgers equation (\ref{Burgers}).

\section*{Acknowledgement}
The author acknowledges a Natural Sciences and Engineering Research Council of Canada (NSERC) postdoctoral fellowship. The author would like to thank its Ph. D. advisor V\'eronique Hussin for helpful discussions.

\end{document}